\documentclass[12pt]{iopart}

\usepackage{color}
\usepackage{graphicx}
\usepackage{cite}
\usepackage{iopams}

\usepackage{multirow}
\usepackage{subfig}

\newcommand{\DWs}{\ensuremath{\Delta W^s}}

\begin{document}

\author{M. Motsch, C. Sommer, M. Zeppenfeld, L.D. van Buuren, P.W.H. Pinkse, and G. Rempe}
\address{Max-Planck-Institut f{\"u}r Quantenoptik, Hans-Kopfermann-Str.\ 1, 85748 Garching, Germany}
\ead{\mailto{pepijn.pinkse@mpq.mpg.de},\mailto{gerhard.rempe@mpq.mpg.de}}

\title{Collisional effects in the formation of cold guided beams of polar molecules}


\begin{abstract}
High fluxes of cold polar molecules are efficiently produced by electric guiding and velocity filtering. Here, we investigate different aspects of the beam formation. Variations of the source parameters such as density and temperature result in characteristic changes in the guided beam. These are observed in the velocity distribution of the guided molecules as well as in the dependence of the signal of guided molecules on the trapping electric field. A model taking into account velocity-dependent collisional losses of cold molecules in the region close to the nozzle accurately reproduces this behavior. This clarifies an open question on the parameter dependence of the detected signal and gives a more detailed understanding of the velocity filtering and guiding process.
\end{abstract}

\maketitle

\section{Introduction and Motivation}
\label{sec:Intro}
Due to their fascinating properties, dense and cold samples of polar molecules offer new perspectives for, e.g., cold and ultracold chemistry, precision measurements, and quantum information processing. Since laser cooling is not applicable to molecules in general, new methods are being developed. For an overview of current research on the production of cold polar molecules as well as possible applications we refer the reader to the present as well as to some past special issues \cite{Doyle2004,Dulieu2006} and reviews on cold polar molecules \cite{Bethlem2003, Krems2005,vandeMeerakker2008}.

In the 1950's Zacharias \emph{et al.} made efforts to produce slow molecules for high precision spectroscopy by directing an effusive molecular beam upwards in a kind of fountain \cite{Zacharias1954,Ramsay:MolecularBeams}. The slow molecules were expected to reverse their travel direction in the gravitational field of the earth already after ascending a small height, such that they could be detected at the base of the fountain. However, no slow molecules could be observed. The explanation for this is twofold: First of all, after leaving the effusive source, the beam dilutes by spreading out so that the density of reversed molecules at the detector is very small. Second, on their way up, the slow molecules are permanently bombarded by fast molecules from behind. In these collisions, the slow molecules gain so much energy and momentum in forward direction that finally the number of slow molecules is so low that none can be detected.

Notwithstanding the experiences with the Zacharias fountain, it is possible to obtain slow molecules by velocity filtering. The filtering efficiency of the slowest molecules from an effusive source can be improved decisively by a number of measures. One is, of course, improved technology in the form of better detectors and a better vacuum. But, more importantly, it is crucial to prevent the molecules from spreading transversely by an appropriate guiding structure. This can be achieved with a magnetic guide for atoms and molecules with an unbound electron \cite{Ghaffari1999,Patterson2007}. For polar molecules, which possess a permanent dipole moment, electric guides are better suited. By providing an electric field which increases in all transverse directions, molecules in low-field-seeking states, i.e., those with a positive Stark shift, will be guided if their transverse velocity is small enough. Moreover, a bend in the guide offers two additional advantages: First, the longitudinal velocity is also limited, since the centripetal force supplied by the electric field only guides molecules around the corner if their longitudinal velocity is small enough. Second, a bend in the guide extracts the slow molecules from the region where collisions with fast molecules from behind are most likely. The molecules are brought rapidly into a ultra-high vacuum region where collisions are rare. These guiding ideas were successfully implemented in our group in the past years to create high-flux beams of slow polar molecules \cite{Rangwala2003,Junglen2004a,Rieger2006,Motsch2008a}.

In this paper we discuss and clarify different aspects of the beam formation in velocity filtering and guiding. For example, the flux of guided molecules can be increased by raising the pressure of the gas in the reservoir and thereby the number of molecules injected into the guide. However, when the reservoir pressure becomes too high, the low-velocity tail of the Boltzmann distribution is depleted by collisions between fast and slow molecules. Apart from the conceptual interest in revisiting the low-velocity tail of the Boltzmann distribution, this study has practical benefits since it explains how the cold polar molecule source can be optimized for a specific application demanding, for instance, molecules with energies below a given trap depth. In \Sref{sec:VelFilter} the necessary theory of the velocity-filtering process is presented.  \Sref{sec:History} reviews some of the experiments performed so far to illustrate the flexibility and applicability of the guide as a source of cold polar molecules. The experimental setup for the guiding experiments is presented in \Sref{sec:Setup}. In \Sref{sec:Meas} we discuss at which parameters collisional effects play a role and how they are observable in the experiment. In \Sref{sec:PScan} the dependence of the signal of guided molecules on the reservoir pressure is discussed which allows first conclusions about effects of collisions on the beam formation. A model is developed for collision-induced losses of slow molecules from the beam, which is discussed in \Sref{sec:Model}. In \Sref{sec:VDep} we show how the model can reproduce the dependence of the signal of guided molecules on the electrode voltage. This allows important conclusions about the detection of the slow molecules with the quadrupole mass spectrometer. Finally, the effect of collisions is confirmed by the observation of shifts in velocity distributions of the guided molecules discussed in \Sref{sec:VelDist}.

\section{Electrostatic velocity filtering of polar molecules}
\label{sec:VelFilter}
Velocity filtering of polar molecules is based on selection of the slowest molecules from a thermal gas. Although the relative fraction of slow molecules with energies below 1\,K is only 10$^{-4}$ for a thermal gas at room temperature according to the Maxwell-Boltzmann velocity distribution, the absolute density of slow molecules can be quite high, when starting with sufficiently high density. The key to velocity filtering is to select the slow molecules in an efficient way, without losing them by collisions with fast molecules, and by accepting a large solid angle.

\begin{figure}
\centering
\includegraphics[width=.6\textwidth]{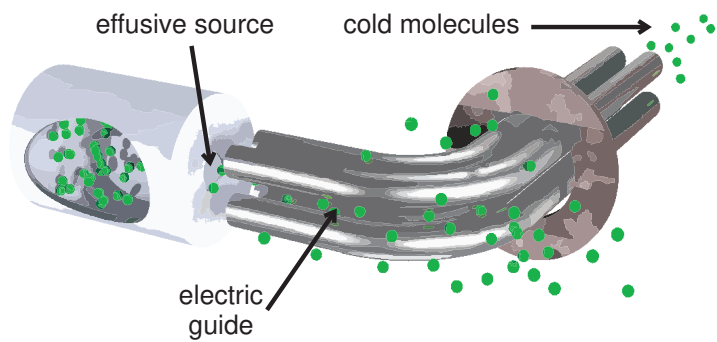}
\caption{\label{fig:idea}Schematic of the electric velocity filtering and guiding experiment. }
\end{figure}

Using suitable electric fields, forces can be exerted on polar molecules. The guiding and trapping potential for slow molecules is realized in the form of an electric quadrupole field, which is created by four high-voltage electrodes. Since electrical fields typically achievable in the laboratory are on the order of 100\,kV/cm, this results in trap depths on the order of a few Kelvin for typical molecular Stark shifts of $\approx$1\,cm$^{-1}$/(100\,kV/cm). Transversely slow molecules in low-field-seeking states are trapped within the region enclosed by high electric fields, while fast molecules escape the guide. Filtering on longitudinal velocity is achieved by bending the guide. When the centrifugal force exceeds the restoring force of the guiding quadrupole field, the fast molecules cannot follow the guide and are lost. How this idea is implemented in the experiment is schematically shown in \Fref{fig:idea}.

The total guided flux of molecules can be calculated from the fraction of molecules in the source which are in a guidable internal state and which have a sufficiently low energy to be trapped in the potential created by the electric fields. In the following, this guided flux is calculated for just one internal molecular state with a fixed Stark shift. The extension to a variety of states is discussed in Ref.~\cite{Motsch2008a}.

\subsection{Source velocity distributions}
Before calculating the flux of guided molecules we have to consider the velocity distributions of molecules in the thermal reservoir from which they are extracted. These are given by a three-dimensional Maxwell-Boltzmann velocity distribution
\begin{equation}
f(v)\:dv = \frac{4}{\sqrt{\pi}\alpha^3} \: v^2 \: \exp(-v^2/\alpha^2) \: dv
\end{equation}
and by a one-dimensional velocity distribution
\begin{equation}
f(v_{x,y,z})\:dv_{x,y,z} = \frac{1}{\sqrt{\pi}\alpha} \: \exp(-v_{x,y,z}^2/\alpha^2) \: dv_{x,y,z}
\end{equation}
with most probable velocity $\alpha = \sqrt{2k_BT/m}$, velocity components $v_i\;(i=x,y,z)$ and total velocity $v=\sqrt{v_x^2+v_y^2+v_z^2}$.

\subsection{Cut-off velocities}
As was already described qualitatively in the introduction into the working principle of the guide, the flux of guided molecules is given by the part of the molecules injected into the guide with velocities below certain transverse and longitudinal cut-off velocities. These cut-off velocities depend on the Stark shift of the molecules and on the properties of the guide. The maximum transverse velocity $v_{max}=\sqrt{2\Delta W^s/m}$ is determined by the Stark shift $\Delta W^s(E_{max})$ reached at the maximum of the trapping field $E_{max}$. If the transverse velocity $v_\bot=\sqrt{v_x^2+v_y^2}$ of the molecule exceeds $v_{max}$, it is lost from the guide. The maximum longitudinal velocity can be obtained by equating the centrifugal force in the bend of radius $R$ and the restoring force of the guide, resulting in $v_{l,max} = \sqrt{\Delta W^s(E_{max}) R/r\,m} = \sqrt{R/2r}\,v_{max}$. Here, $r$ is the free inner radius of the guide, at which the maximum trapping field is reached. As almost every particle is guided if $v_l < v_{l,max}$ and $v_\bot < v_{max}$, this results in higher efficiencies as compared to filtering by, e.g., rotating filter wheels and apertures.

\subsection{Flux of guided molecules}
To calculate the flux of guided molecules, the velocity distributions of molecules injected into the guide are integrated to the maximum trappable velocity. In the limit of small cut-off velocities $v_{max}$ and $v_{l,max}$ compared to the thermal velocity $\alpha$, the exponential $\exp(-v^2/\alpha^2)$ in the thermal velocity distributions can be replaced by 1. The guided flux $\Phi$ of a molecular state with a Stark energy $\Delta W^s$ is then given by
\begin{equation}
\begin{array}{ll}
\label{eq:VelFilteringFlux}
\Phi &= \int\limits_{v_x=-v_{max}}^{v_{max}} \int\limits_{v_y=-v_{max}}^{v_{max}} \int\limits_{v_z=0}^{v_{l,max}} f(v_x) f(v_y) \;v_z\:f(v_z)\; dv_x dv_y dv_z\\
&\propto v_{max}^4 \propto (\Delta W^s)^2,
\end{array}
\end{equation}
where the dependence of the transverse cut-off velocity on the longitudinal velocity is neglected, and $\Phi$ is normalized to the flux of molecules out of the nozzle. The guided flux can hence be described by a function $f$ which gives the fraction of guidable molecules for a given electric field, $\Phi=f(\Delta W^s) \propto (\Delta W^s)^2$. Throughout the rest of this paper we assume linear Stark shifts $\Delta W^s \propto E_{max} \propto U$ which is appropriate for ammonia molecules in the investigated range of trapping electric fields.

\section{History, developments and extensions of the electric guide}
\label{sec:History}

Velocity filtering using an electrostatic quadrupole guide was first demonstrated with formaldehyde (H$_2$CO) and deuterated ammonia (ND$_3$) molecules \cite{Rangwala2003,Junglen2004a}. There, fluxes of 10$^{10}$--10$^{11}$ molecules/s at peak densities of 10$^9$\,cm$^{-3}$ were achieved using typical laboratory guiding fields of 100\,kV/cm. The molecules in these guided beams had typical velocities of 50\,m/s, corresponding to a translational temperature of a few Kelvin. Since electrostatic fields were used in these experiments, only molecules in low-field-seeking states could be guided.

To extend the velocity-filtering technique to molecules in high-field-seeking states, a dynamical trapping potential must be created. To accomplish this, the electric guide is switched between two dipolar configurations in a periodic manner. Thereby, molecules in low-field-seeking and high-field-seeking states alike can simultaneously be trapped in a time-averaged potential. In the experiment, it was shown that guiding of polar molecules in such alternating electric fields is indeed possible \cite{Junglen2004}. However, simulations suggested that only low-field-seeking molecules reached the detection volume at the exit of the guide.
To show that high-field-seekers can indeed be trapped by alternating fields, an electric trap for neutral rubidium atoms was set up \cite{Rieger2007}. Another motivation for this experiment was the prospect of trapping atoms and molecules simultaneously in the same trap to achieve, e.g., sympathetic cooling of the molecules. To trap them, rubidium atoms are pre-cooled in a magneto optical trap and then magnetically transferred to the all-electric trap. For these high-field-seeking rubidium atoms storage times of a few hundred milliseconds in the electric trap were demonstrated.

Since many experiments benefit from long interaction times between the species of interest and, e.g., laser fields or other species for the study of collisions, the electrostatic quadrupole guide was connected to an electrostatic trap for molecules \cite{Rieger2005}. The trap consists of several ring electrodes which match in a natural way to the electric quadrupole guide. With this continuously loadable trap, storage times of 130\,ms at temperatures of 300\,mK were demonstrated for ammonia molecules, limited by the size of the trap. This opens the perspective of reaching longer times by a more elaborate design of an electrostatic trap for polar molecules.

In the first guiding experiments molecules such as formaldehyde or ammonia were used. A polar molecule which is of highest interest to many different fields is water. However, due to its molecular structure, ordinary (H$_2$O) or fully deuterated (D$_2$O) water exhibits quadratic Stark shifts. Therefore, guiding and trapping is significantly more challenging than with formaldehyde or ammonia which exhibit large, linear Stark shifts . Nevertheless, with the electrostatic guiding and velocity filtering technique cold guided beams of deuterated water (D$_2$O) could be produced, showing the flexibility of this method \cite{Rieger2006}.
Recently, these experiments with cold, guided water beams were extended to include all the water isotopologues H$_2$O, D$_2$O and HDO \cite{Motsch2008a}. Although seeming very similar at a first glance, these molecules show a totally different behavior when exposed to the guiding electric fields. This allowed to investigate different aspects of the velocity-filtering process such as its dependence on rotational states of the polar molecules.

The flux of cold guided polar molecules is so high that it can easily be detected with a quadrupole mass spectrometer (QMS). This has the additional advantage of being robust and flexible. However, no direct information on the population of internal states of the molecules is available. Therefore, the detection with the QMS was combined with a laser spectroscopy technique \cite{Motsch2007}. In the guide, formaldehyde (H$_2$CO) molecules were optically pumped to an excited state which dissociates rapidly. If this optical pumping is done in a state-dependent way, the laser-frequency-dependent decrease of the QMS signal allows to infer information on populations of individual rotational states of the guided molecules. In principle this technique is applicable to other species as well, since coupling to an unguided state is sufficient to generate losses from the guided population.

In all experiments mentioned so far the molecules were extracted from a thermal reservoir which could, at most, be cooled to 150\,K. The source would, however, benefit a lot from starting with an ensemble already at cryogenic temperatures. Therefore, electrostatic velocity filtering was recently combined with buffer-gas cooling \cite{Weinstein1998,Patterson2007,vanBuuren2008,Sommer2008}. Here, molecules are injected into a cryogenic helium buffer gas through a heated input capillary. The molecules thermalize by collisions with the cold helium atoms, and finally leave the buffer-gas cell through an exit aperture. This beam leaving the cell is then collected by the electric guide. With this setup, we produce guided beams with densities of 10$^9$\,cm$^{-3}$ and fluxes of 10$^{11}$\,molecules/s at velocities of 60\,m/s were produced. By collisions with the cold helium buffer gas not only the translational degrees of freedom, but also internal excitations are cooled. With the depletion spectroscopy technique for formaldehyde it could be verified that the rotational degrees of freedom are cooled down to the temperature of the buffer gas, 5\,K, resulting in a state-selected beam with at least 80\,\% of the population in a single rotational state.

To summarize, by velocity filtering continuous guided beams of polar molecules with high fluxes of 10$^{11}$\,molecules/s at densities of 10$^9$\,cm$^{-3}$ can be produced. The technique is very flexible and can easily be adopted to different species. By combination with cryogenic buffer-gas cooling the molecules are also cooled internally, such that a high-flux, state selected, continuous guided beam of polar molecules is at hand. The flux of molecules resulting from the cryogenic buffer-gas source is, however, similar to the one obtained with a room-temperature reservoir. The reason for this are collisions of the slow molecules with helium atoms when leaving the buffer-gas cell. Similar effects can also occur when a thermal source is used. However, here the slow molecules collide with fast molecules streaming out of the nozzle. After a brief discussion of the experimental setup, we present a detailed analysis of these collisional effects on the beam formation process employing a room-temperature source.

\section{Experimental setup}
\label{sec:Setup}

\begin{figure}
\centering
\includegraphics[width=.6\textwidth]{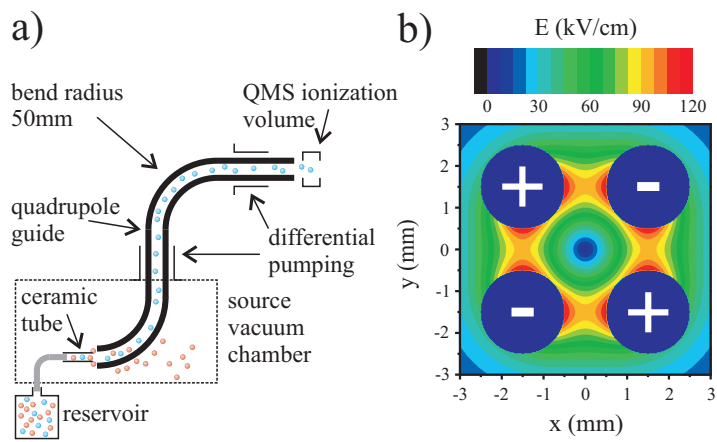}
\caption{\label{fig:ExpSetup}Experimental setup. a) Molecules from the thermal reservoir enter the electric guide through the ceramic nozzle. Slow molecules are trapped in the quadrupole field and guided to an ultra-high-vacuum chamber, where they are detected by the quadrupole mass spectrometer. b) Electric field distribution in the quadrupole guide for $\pm$5\,kV electrode voltage, resulting in a trapping electric field exceeding 93\,kV/cm.}
\end{figure}

The experimental setup used for the experiments is the same as the one described in \cite{Motsch2008a} and is shown schematically in \Fref{fig:ExpSetup}. The quadrupole guide for polar molecules is located in three interconnected ultra-high vacuum chambers. Molecules are injected into the guide through a ceramic tube ($\varnothing$ 1.5\,mm, length 9.5\,mm) which is connected to a liquid nitrogen reservoir and equipped with a heater element, such that its temperature can be adjusted in the range 100--400\,K\@. To ensure thermalization the Teflon tube for the molecular gas passes a meandering cooling stage of 0.3\,m length before entering the ceramic tube. Through the rest of the paper we shortly call this ceramic tube with the attached cooling stage "nozzle". By cooling the nozzle, the fraction of slow molecules increases, as the Maxwell-Boltzmann velocity distribution gets compressed and shifted to lower velocities. This increases the guided flux. Furthermore, the number of thermally occupied states is reduced, leading to an improved purity of the guided beam as shown by depletion spectroscopy \cite{Motsch2007}. The nozzle is located in a vacuum chamber, in which a base pressure of 10$^{-9}$\,mbar is achieved by a 500\,l/s turbo molecular pump. When flowing gas, the pressure in the chamber rises to typical values of a few 10$^{-7}$\,mbar. From this pressure rise and the known pumping speed the gas flow rate through the nozzle can be determined.

In contrast to previous experiments, where a bend radius of 25\,mm \cite{Rangwala2003,Junglen2004a} or only 12.5\,mm \cite{Rieger2005} was used, the current setup employs a larger bend radius of 50\,mm. This has the advantage of an increased flux, which is especially valuable when looking for small signals as, for instance, for molecules with quadratic Stark shifts such as H$_2$O or D$_2$O \cite{Motsch2008a}. The increase in flux is also very valuable when doing differential measurements such as in depletion spectroscopy \cite{Motsch2007,vanBuuren2008}, or when performing experiments for small reservoir pressures as the ones described in this paper. The molecules are guided around two bends and through two differential pumping stages to an ultrahigh vacuum chamber, where they are finally detected by a quadrupole mass spectrometer (QMS, Pfeiffer QMG422). In the QMS, the guided molecules are ionized by electron impact in a cross-beam geometry. The ions are then mass filtered, before in the final stage single ion counting using a secondary electron multiplier is performed. The electronic signals are amplified, shaped to TTL pulses, passed through an isolating amplifier, and are finally recorded using a multi-channel scalar card. This detection of the guided molecules by the QMS is not sensitive to the internal states. However, by combination with ultraviolet laser spectroscopy in the guide, state-sensitive detection can be achieved \cite{Motsch2007}.

\section{Measurements of collisional effects}
\label{sec:Meas}

For the description of the velocity-filtering process in \Sref{sec:VelFilter}, a purely effusive source was assumed. This means that the velocity distribution of the molecules injected into the guide is assumed to directly reflect the thermal velocity distribution of molecules in the nozzle. As soon as collisions between molecules come into play, i.e.\ the mean free path $\Lambda$ of the molecules becomes comparable to the dimensions $d$ of the nozzle $\mathcal{O}(\Lambda)\approx\mathcal{O}(d)$, this condition is no longer fulfilled.

From an experimental point of view it is therefore instructive to perform measurements for varying inlet pressures. Thereby, effects caused by collisions of molecules in the nozzle or in the higher-density region directly behind the nozzle can be investigated. Furthermore, such an experiment is also interesting for more practical reasons. When using the guide as a source for cold polar molecules one wants to operate the system in a parameter regime which results in the maximum flux of guided molecules. This optimum value depends on the requirements of the specific application. When a high flux of molecules with velocities of a few 10\,m/s is needed, collisions removing predominantly the slowest molecules are of no concern. Therefore, a higher reservoir pressure can be chosen, resulting in a higher flux of molecules in the guide. This changes if one is interested in only the slowest molecules. Then, the slow molecules in the beam emerging from the nozzle must not be removed by any means, suggesting the use of a smaller reservoir pressure to reduce collisional effects.

To investigate these effects, the reservoir pressure is varied between 0.01\,mbar and 2.0\,mbar, as measured by the Pirani pressure gauge used for pressure regulation. These pressures span a wide range around the reservoir pressure of 0.1\,mbar used for most other experiments so far \cite{Motsch2008a}. To make this numbers comparable to other guiding experiments with similar setups, the gas flow through the nozzle is calculated from the pumping speed and the observed pressure rise in the source chamber. For a reservoir pressure of 0.1\,mbar we determine a gas flow rate of $1\times10^{-4}$\,mbar$\cdot$l/s with the nozzle assembly at room temperature.

Due to this large variation in pressure, count rates of the QMS also vary over a large range. By measurements with different emission currents of the QMS ionization unit it was assured that the observed effects are indeed caused by differences in the molecular flux and not related to saturation of the ionization process or to nonlinearities in the detection process. All data used throughout the rest of this paper was taken with 0.10\,mA emission current, except for the velocity distributions measured at $\pm$3\,kV and 0.01\,mbar reservoir pressure. Here, the emission current was increased to 0.20\,mA to reduce the necessary measurement times. By comparing velocity distributions measured at higher reservoir pressures for different emission currents it was made sure that this 0.20\,mA measurement does not introduce systematic shifts.

The experiments are performed as a series of time of flight (TOF) measurements, where we switch the high voltage (HV) which produces the guiding electric fields on and off in a fixed timing sequence. To subtract contributions of background gas to the QMS signal, the difference in the steady state QMS signal with HV applied to the guide and HV switched off is used to determine the signal of guided molecules.

\section{Pressure dependence of velocity filtering}
\label{sec:PScan}

\begin{figure}
\centering
\includegraphics[width=.6\textwidth]{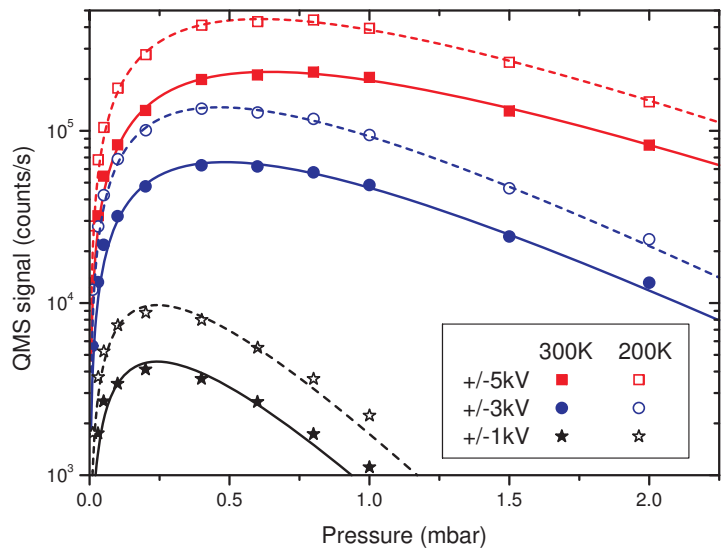}
\caption{\label{fig:PressScan}Signal of guided molecules as a function of reservoir pressure, measured for different combinations of electrode voltage and source temperatures. The curves are fits to the data based on the model for the beam formation.}
\end{figure}

\Fref{fig:PressScan} shows the signal of guided molecules as a function of reservoir pressure. In this plot, measurements for different electrode voltages and source temperatures are combined. The curves are fits to the data based on the model for the beam formation which will be presented in \Sref{sec:Model}. The model very well reproduces the shape of the pressure dependence for all electrode voltages. Also, it shows the good sensitivity of our method (guiding at only $\pm$1\,kV) with count rates of  $10^3$--$10^5$\,counts/s.

The first feature to observe is that, starting at zero pressure, the signal of guided molecules increases with reservoir pressure, reaches a maximum, and finally drops again. This behavior can be expected considering the working principle of the guide described in \Sref{sec:VelFilter}. The guide accepts molecules up to certain cut-off velocities, which depend on the Stark shift and the applied trapping field. For small pressures, an increase in pressure increases the number of molecules injected into the guide while the effusive character of the source is maintained. In contrast to this, for high pressures collisions in the nozzle and in the region behind the nozzle become important.  Thereby, the slow molecules are eliminated from the beam. This leads to a decay of the signal of guided molecules. The maximum is reached at a point where these two effects are balanced: Although more molecules are injected into the guide, the number of molecules with guidable velocities stays constant due to the diminishing fraction of slow molecules.

A second feature to observe is the increase of the signal of guided molecules with applied electrode voltage \cite{Rangwala2003,Junglen2004a,Motsch2008a}. As shown in \Sref{sec:VelFilter}, the cut-off velocities get higher when the electrode voltage is raised. For fixed reservoir pressure, this increases the fraction of molecules with guidable energies. The dependence of the guided signal as a function of electrode voltage will be discussed in more detail in \Sref{sec:VDep}, since it allows to infer many details of the beam formation. By comparing measurements for fixed electrode voltage, it can be seen that cooling of the nozzle increases the signal of guided molecules \cite{Junglen2004a}. By cooling the source from 300\,K to 200\,K the signal of guided molecules increases by a factor 2.2.

\begin{figure}
\centering
\includegraphics[width=.6\textwidth]{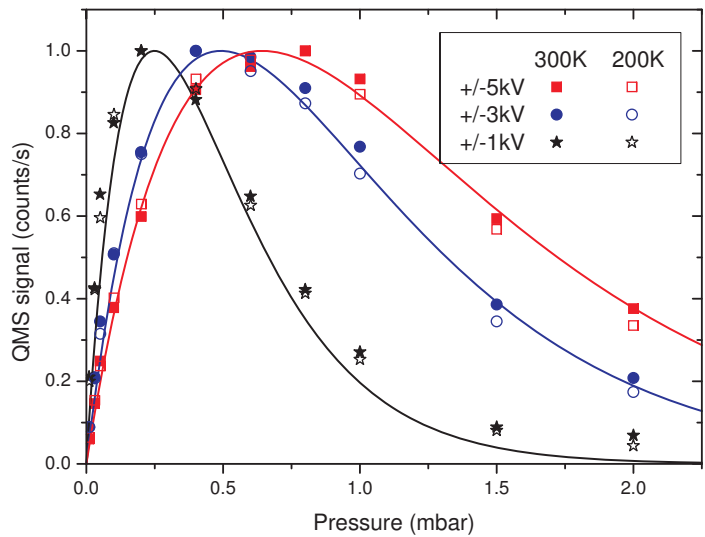}
\caption{\label{fig:NormPressScan}Signal of guided molecules as a function of reservoir pressure, measured for different combinations of electrode voltage and source temperatures. The data for each electrode voltage is individually normalized. The curves are fits to the data based on the model for the beam formation. For the $\pm$1\,kV measurements the data at reservoir pressure $\geq$1\,mbar is influenced by the rise of the background pressure in the detection chamber.}
\end{figure}

\Fref{fig:NormPressScan} shows the normalized guided signal as a function of reservoir pressure. This figure allows to maximize the flux from the electric guide when it is used as a source for other experiments. By choosing the optimum pressure setting, the number of guided molecules for a given electrode voltage, i.e.\ up to a certain cut-off velocity, is maximized. From this figure it can also be seen more clearly that the maximum of the curve shifts towards larger pressures when higher electrode voltages are used.

This allows to infer the main feature of the model used to describe the beam formation. For higher electrode voltages, faster molecules can be guided. The increase in optimum pressure with applied electrode voltage shows that for these faster molecules collisions start playing a role only at higher pressures and thereby higher densities. This suggests that the collision probability scales with the time spent in the high pressure region or, equivalently, with $1/v$, where $v$ is the molecule velocity. In \Sref{sec:Model}, this assumption will be used as the basis of our model for the beam formation.

Comparing the 300\,K and 200\,K measurements it should be noted that not only the shape of the curves but also the optimum pressures for a given electrode voltage are identical. Note that the line conductance is dominated by parts for which the temperature is not varied. Hence, the line conductance and the gas flow through the nozzle for a given reservoir pressure does not depend on the temperature. The probability for a slow molecule to undergo a collision in the "high pressure region" stays constant, since the same flux of molecules is streaming out of the nozzle. Therefore one observes the same dependence on reservoir pressure, independent of the nozzle temperature.

\section{Model of velocity filtering including collisional losses}
\label{sec:Model}

The effect of collisions is included into the theory of velocity filtering presented in \Sref{sec:VelFilter} as follows. The probability for a slow, in principle guidable molecule to undergo a collision with some fast molecule is proportional to the time $t_{c}$ which the slow molecule spends in the region where collisions are most probable. In a collision the velocity of a slow molecule is likely to increase, such that it cannot be guided anymore. Hence, the probability for a slow molecule to be lost from the guided beam will be inversely proportional to the molecule's longitudinal velocity $v_z$ due to $t_{c} \propto 1/v_{z}$. Note that only the velocity of the molecule in longitudinal direction determines the time spent in the high-pressure region, since no pressure variation is expected over the transverse extent of the nozzle opening which is matched to the dimensions of the guide. To account for this velocity-dependent loss, an additional factor $\exp{(-b/v_z)}$ is included in the longitudinal velocity distribution. Here, $b$ is a density dependent "boosting" parameter. Putting this together with the original thermal velocity distribution in the reservoir, the longitudinal velocity distribution $f_b(v_z)$ of molecules injected into the guide is given by
\begin{equation}
f_b(v_z)\:dv_z = \frac{1}{N} \: \exp(-b/v_z) \: \exp(-v_z^2/\alpha^2) \: dv_z,
\end{equation}
where $N$ is a constant such that the total flux coming out of the nozzle is normalized to $\int_{0}^{\infty}f_b(v_z)\:dv_z=1$.
\begin{figure}
\centering
\includegraphics[width=0.99\textwidth]{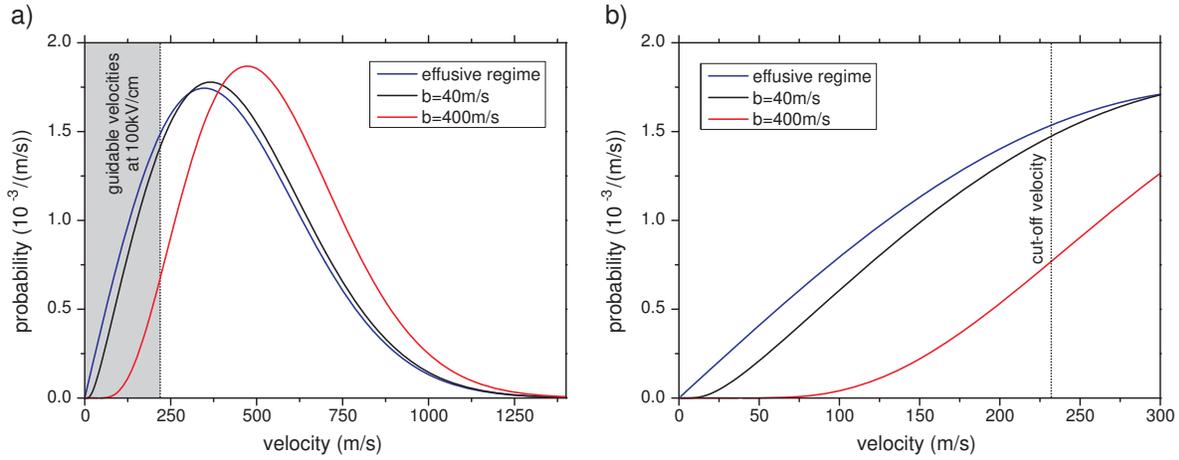}
\caption{\label{fig:BoostVelDist} a) Longitudinal velocity distributions of molecules emerging from the thermal source including the boosting term. Shown are distributions for different values of the boosting parameter $b$, corresponding to reservoir pressures of 0.1\,mbar ($b$=40\,m/s) and 1.0\,mbar ($b$=400\,m/s) in the experiment. Guidable velocities are indicated for a trapping electric field of 100\,kV/cm and a Stark shift of $\Delta$W$^s$=2\,cm$^{-1}$, resulting in a longitudinal cut-off velocity of 230\,m/s for ND$_3$. b) is a zoom into the region of guidable velocities.}
\end{figure}
This longitudinal velocity distribution modified by collisions is shown in \Fref{fig:BoostVelDist} for different values of the parameter $b$. It can be seen that the fraction of slow molecules gets more and more reduced for increasing collision probability. \Fref{fig:BoostVelDist} also shows that the mean velocity of the molecules coming out of the nozzle increases with $b$. Therefore, we shortly speak of a \emph{boosting} of the beam. This effect is observable in experiments with small electrode voltages. At these settings only the slowest molecules which are most affected by the boosting are accepted by the guide, which affects the measured signals strongly. Boosting can also be observed for higher electrode voltages, when the reservoir pressure is increased. This results in larger boosting effects, which influence also faster molecules.

The flux of guided molecules $\Phi(U)$ as a function of the electrode voltage $U$ can be calculated by integration over the velocity distributions up to the longitudinal and transverse cut-off velocities as described in \Sref{sec:VelFilter}. We again neglect the factor $\exp(-v^2/\alpha^2)$, since $v\ll\alpha$. This results in
\begin{equation}
\begin{array}{ll}
\Phi(U) & = \int\limits_{v_x=-v_{max}}^{v_{max}} \int\limits_{v_y=-v_{max}}^{v_{max}} \int\limits_{v_z=0}^{v_{l,max}} f(v_x) f(v_y) \; v_z \: f_b(v_z)\; dv_x dv_y dv_z\\
& \propto v_{max}^2 \int \limits_{v_z=0}^{v_{l,max}} v_z \: \exp(-b/v_z) \;dv_z
\propto U \int \limits_{v_z=0}^{v_{l,max}}
v_z\:\exp(-b/v_z) \;dv_z,
\end{array}
\end{equation}
where use has been made of $v_{max},v_{l,max}\propto \sqrt{\DWs}$ and $\DWs \propto U$.

In the experiments, however, the QMS measures density. Since the density $n$ is connected to the flux by $n=\Phi/v$, the signal of guided molecules $S(U)$ is expected to be given by
\begin{equation}
\label{eq:BoostedSignal}
S(U) \propto U \int \limits_{v_z=0}^{v_{l,max}}\exp(-b/v_z)\;dv_z.
\end{equation}
Without any boosting, i.e.\ assuming a perfectly effusive source ($b=0$), this results in $S(U)\propto U^{3/2}$. Note that in previous experiments a dependence $S(U)\propto U^2$ was observed, which seemed to indicate a flux measurement \cite{Rangwala2003,Junglen2004a,Rieger2006,Motsch2008a}.

\section{Electrode voltage dependence of velocity filtering}
\label{sec:VDep}

\begin{figure}
\centering
\subfloat{
\includegraphics[width=0.6\textwidth]{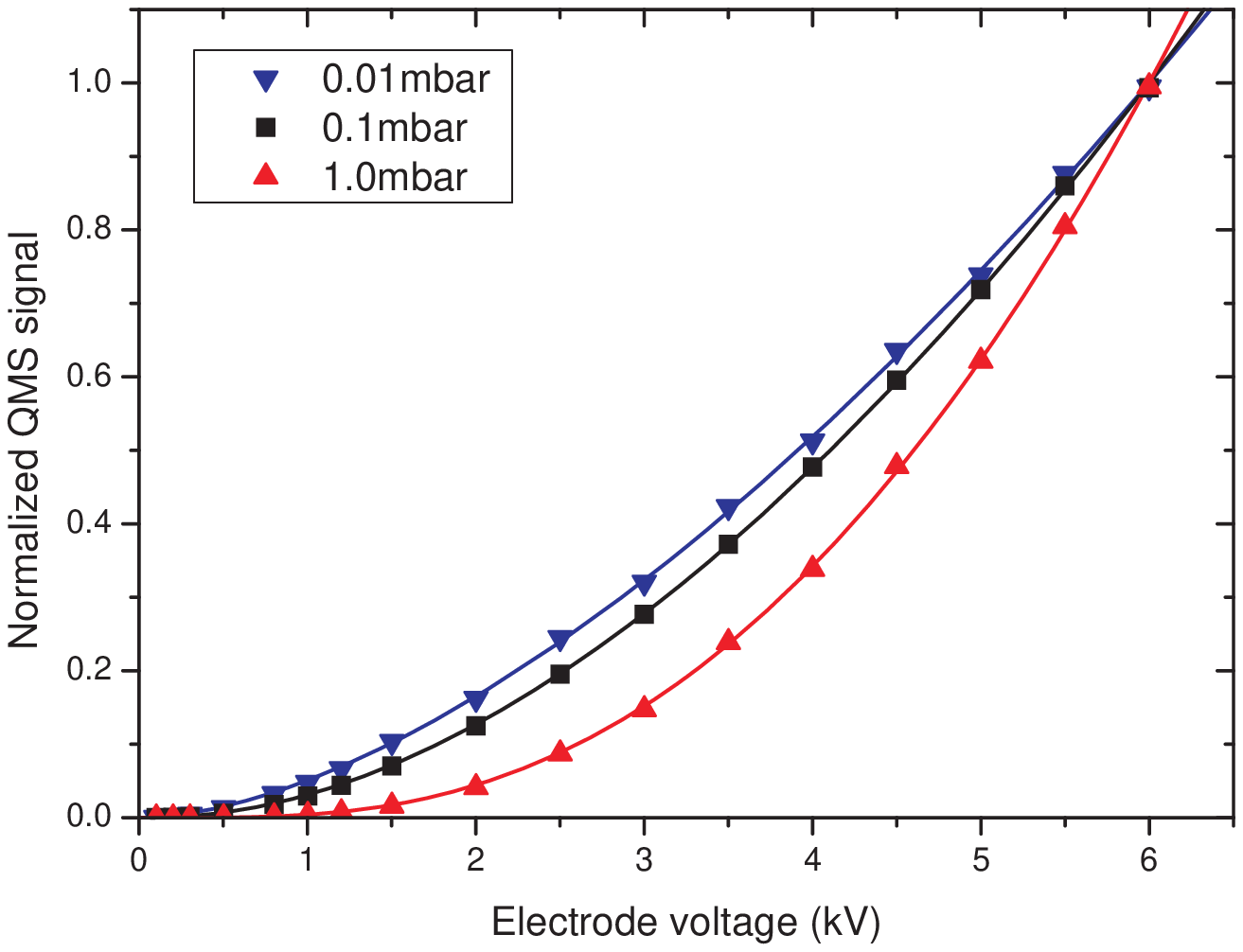}
}
\\
\subfloat{
\includegraphics[width=0.6\textwidth]{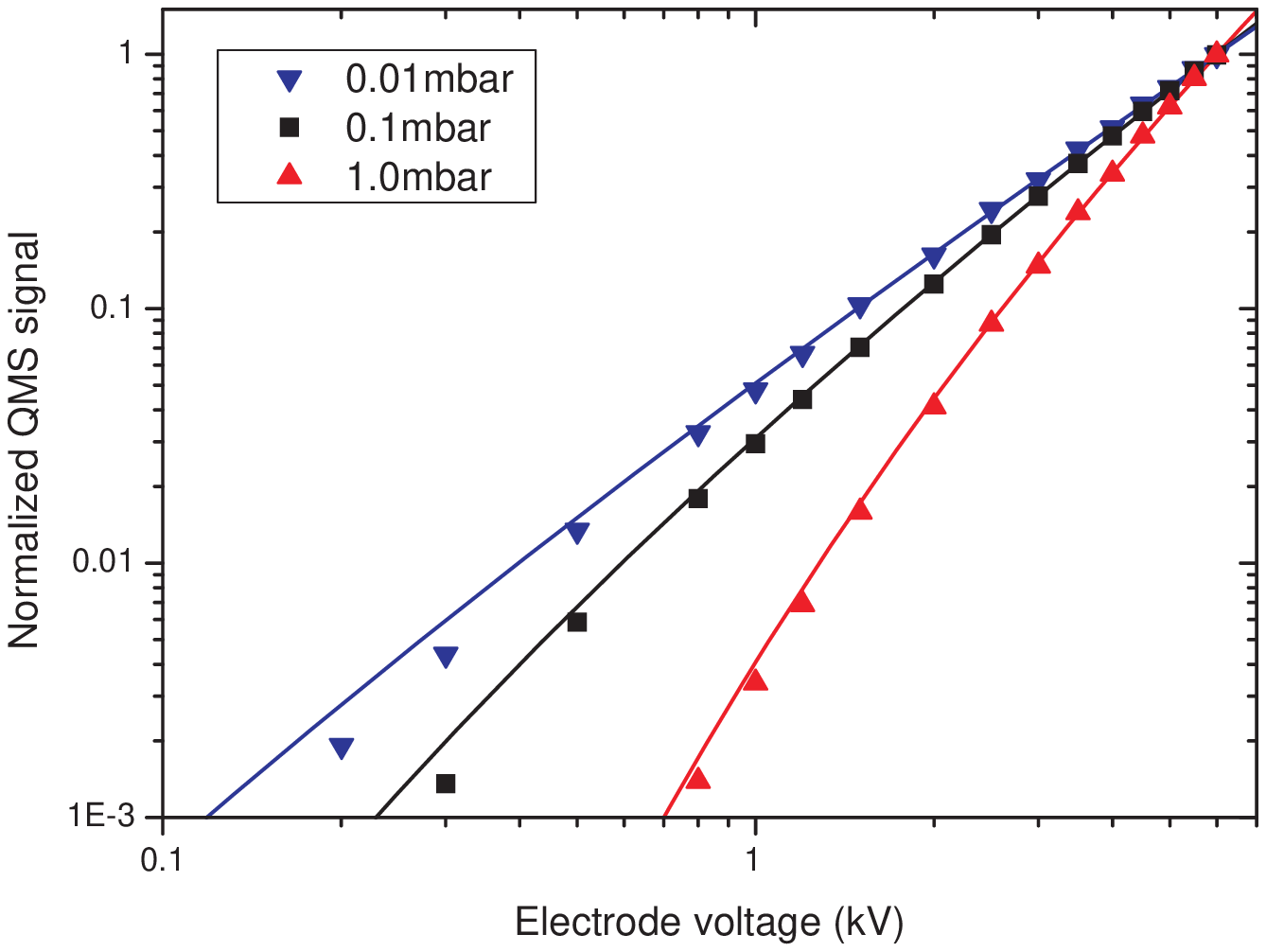}
}
\caption{\label{fig:VoltDep}Dependence of the signal of guided molecules on the electrode voltage. The curves are fits using the model for the beam formation including collisions in the high pressure region. The lower plot shows the same data and fit curves with logarithmic axes.}
\end{figure}

\Fref{fig:VoltDep} shows the signal of guided molecules as a function of the applied electrode voltage for different settings of the reservoir pressure. The solid curves are fits using \eref{eq:BoostedSignal} which takes into account collisional depletion of slow molecules from the guided beam. The data of the individual voltage scans is normalized to the fit value at 6kV electrode voltage. From this figure, several features can be observed. Increasing the reservoir pressure results in more collisions and a larger boosting parameter.  Therefore, a larger fraction of molecules is depleted, causing a relatively smaller signal at lower electrode voltages where only the slowest molecules are guided. This causes the curve in the lin-lin plot to bend stronger. From the log-log plot the different character of the voltage dependencies can clearly be observed. For electrode voltages above 1 or 2\,kV  the fit curves for the 0.01\,mbar and 0.1\,mbar measurements are nearly straight lines in the log-log plot. However, the slope of the fit curve changes with reservoir pressure, indicating different power laws. The deviations for small electrode voltages may be explained by the behaviour of the Stark shift: For small trapping electric fields the Stark energy of the ND$_3$ molecules becomes comparable to the inversion splitting, which leads to quadratic Stark shifts and therefore a smaller fraction of guided molecules.

The data shown in \Fref{fig:VoltDep} are taken at 300\,K source temperature. For a source temperature of 200\,K exactly the same behavior is observed, as already found in the pressure dependence study discussed in \Sref{sec:PScan}. Since the flux of molecules out of the nozzle does not change with temperature in our setup, the collision probability for a slow molecule shows the same dependence on the reservoir pressure, independent of the source temperature.

\begin{figure}
\centering
\includegraphics[width=0.6\textwidth]{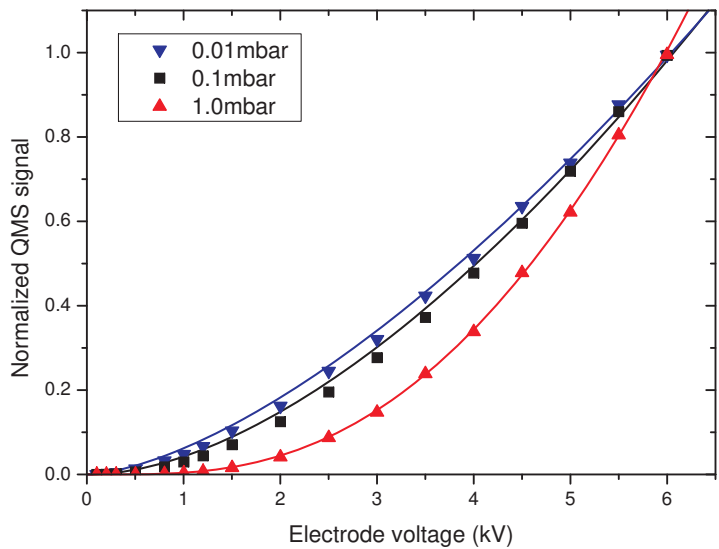}
\caption{\label{fig:VoltDepScaled}Dependence of the signal of guided molecule on the electrode voltage. The data at 1.0\,mbar has been fitted with the model for the beam formation and serves a reference. For the other curves the boosting parameter $b$ is scaled linearly with pressure.}
\end{figure}

The boosting parameter $b$ is expected to depend linearly on the flux of molecules out of the nozzle, i.e.\ in the collision region, and thereby also on the reservoir pressure. We therefore use the measurement at high reservoir pressure of 1.0\,mbar as a reference. Here, also relatively fast molecules are affected by the boosting. As a result, the boosting can be observed at higher electrode voltages, where other systematic effects are smaller. We then scale the boosting parameter linearly with reservoir pressure. \Fref{fig:VoltDepScaled} shows the electrode voltage dependence of the signal of guided molecules together with curves of the model \eref{eq:BoostedSignal} using scaled values of the boosting parameter $b$. At large electrode voltages the data is well described by these curves. For smaller electrode voltages other effects which were discussed above come into play and reduce the signal of guided molecules as compared to the predictions made by this model.

\begin{figure}
\centering
\subfloat[0.01\,mbar reservoir pressure]{
\begin{minipage}{0.98\textwidth}
\includegraphics[width=0.47\textwidth]{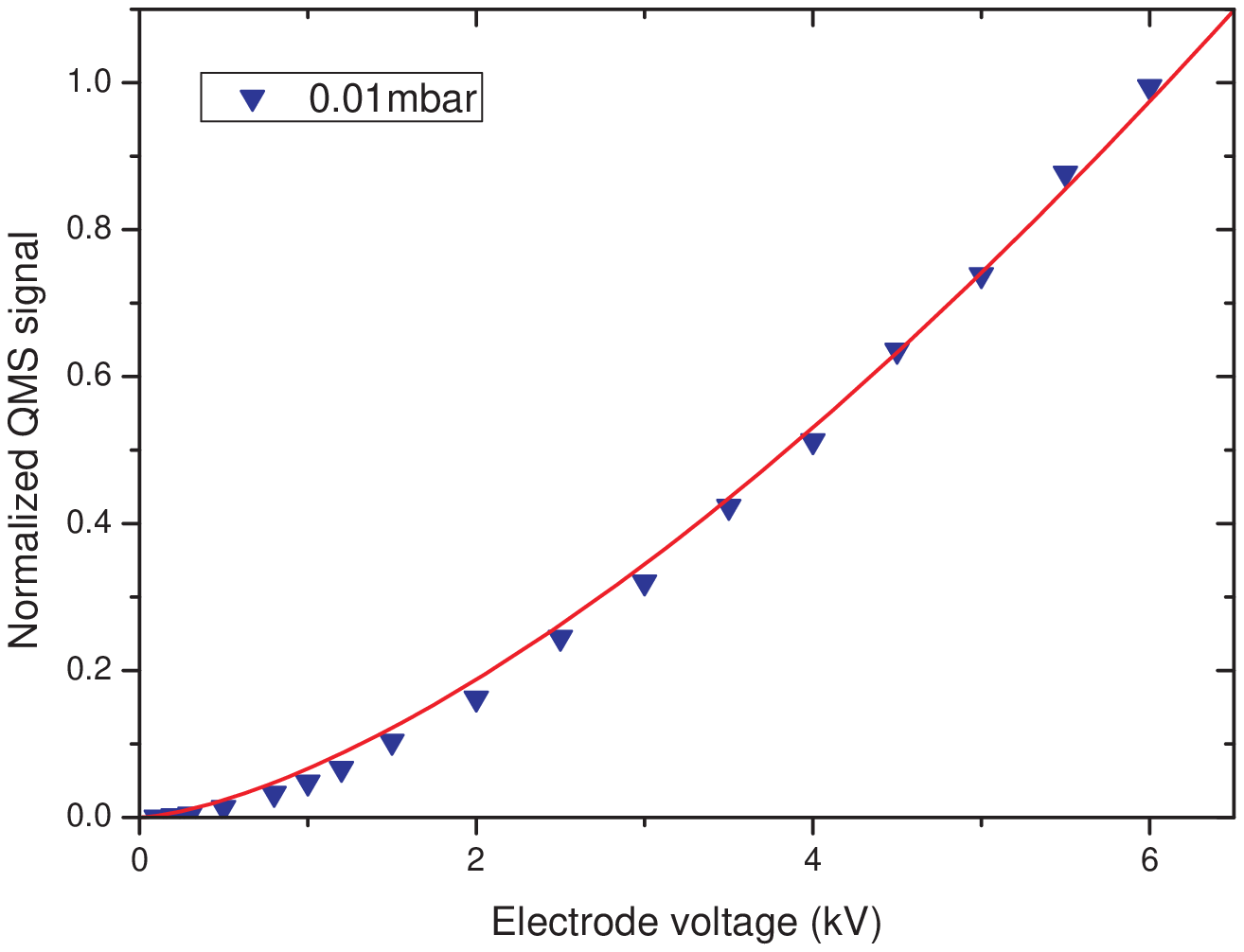}
\hfill
\includegraphics[width=0.47\textwidth]{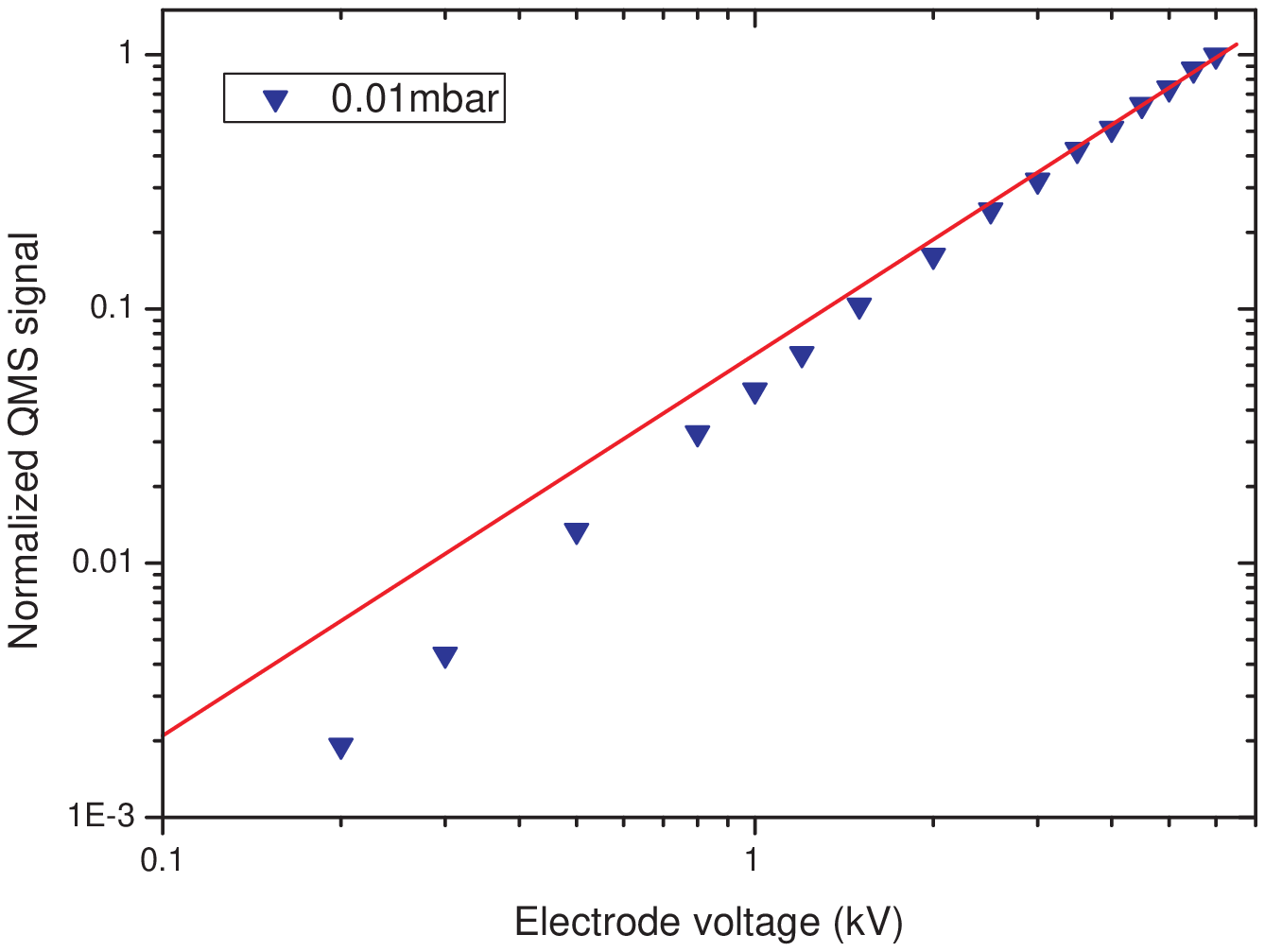}
\end{minipage}
}
\\
\subfloat[0.1\,mbar reservoir pressure]{
\begin{minipage}{0.98\textwidth}
\includegraphics[width=0.47\textwidth]{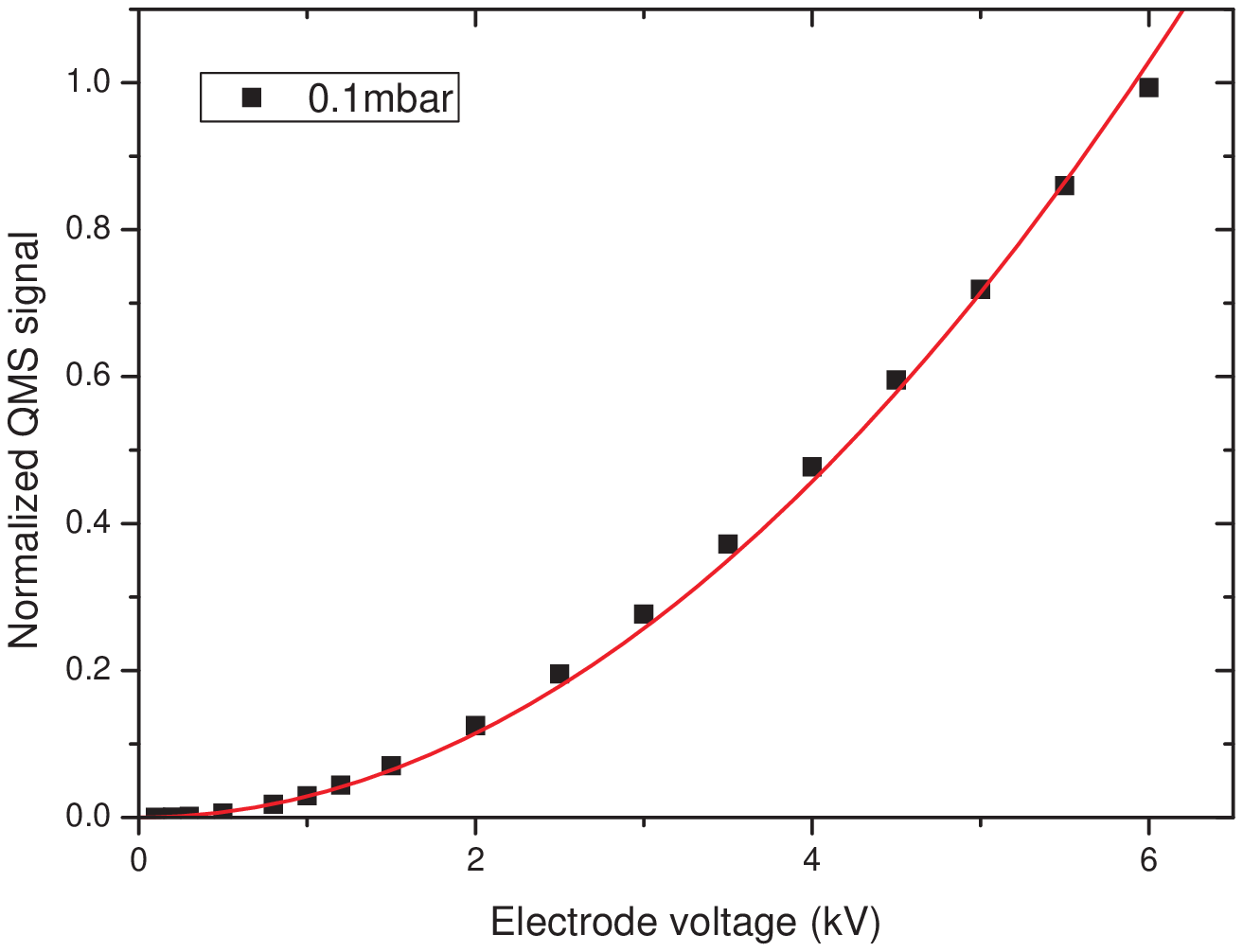}
\hfill
\includegraphics[width=0.47\textwidth]{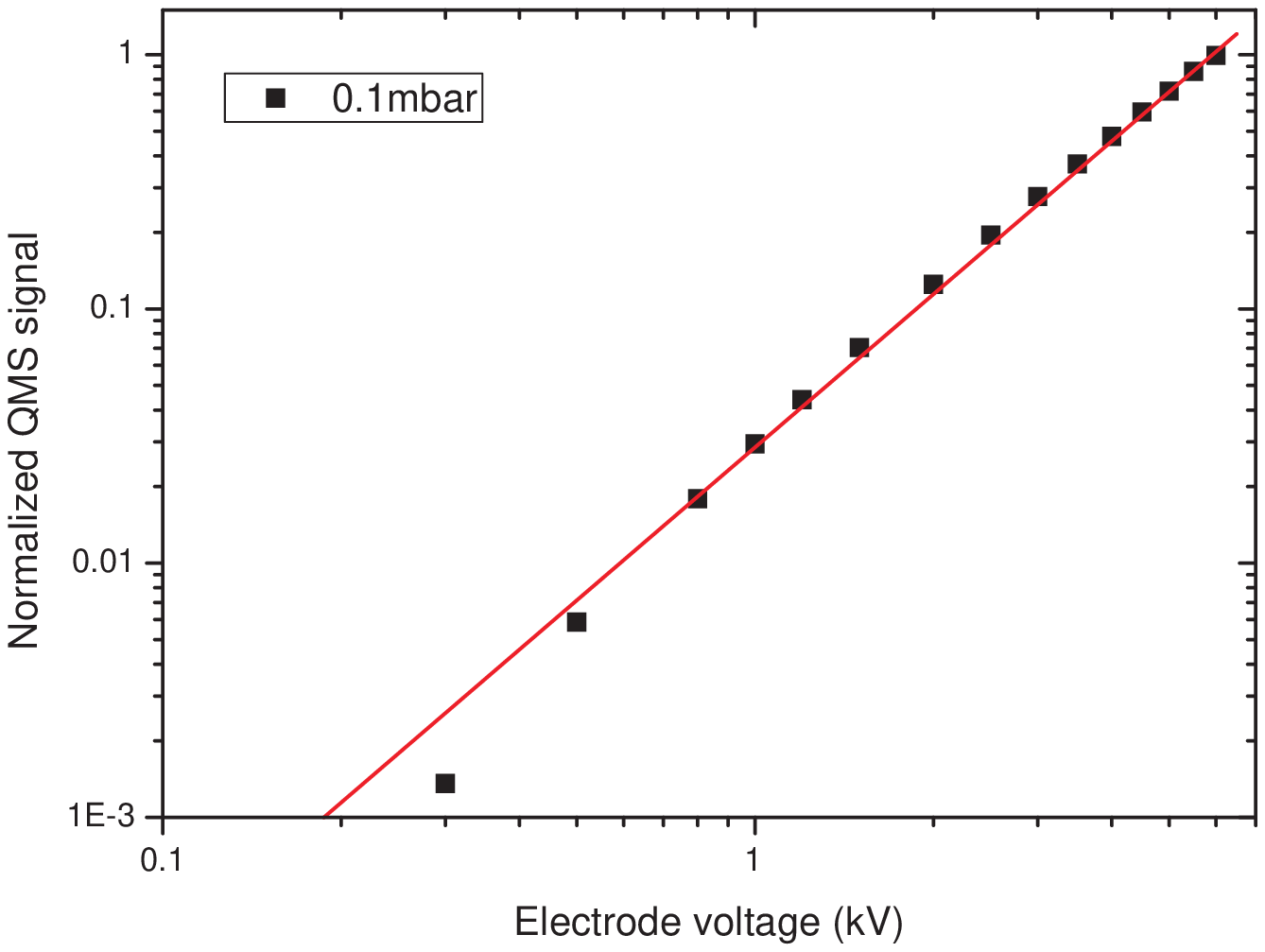}
\end{minipage}
}
\caption{\label{fig:VoltDepPoly}Dependence of the signal of guided molecules on the electrode voltage. The curves are fits using (a) $U^{3/2}$ and (b) $U^2$.}
\end{figure}

To connect the model for the guided molecule signal including collisions in the high-pressure region to previous experiments we fit our data with a polynomial. \Fref{fig:VoltDepPoly}~a) shows the dependence of the guided molecule signal for very low reservoir pressure (0.01\,mbar). For voltages above 1--2\,kV the data is well described by a $U^{3/2}$ dependence represented by the solid line. This is exactly the electrode-voltage dependence expected for an ideal effusive source, as discussed in \Sref{sec:Model}.

\Fref{fig:VoltDepPoly}~b) shows the signal of guided molecules in a medium pressure (0.1mbar) regime. For most experiments utilizing the guide, the source is operated in this range of pressures \cite{Junglen2004a,Motsch2008a}. Here, the data is shown together with a fit of a $U^2$ electrode voltage dependence. In earlier experiments \cite{Rangwala2003,Junglen2004a,Rieger2006} the good agreement between the experimental data and the $U^2$ model suggested a flux measurement in combination with an ideal effusive source. However, calibrations of fluxes and densities of guided molecules correctly assumed a QMS signal proportional to the density in the ionization volume. Now, the setup of a new and improved electric guide used for the experiments presented in this paper allowed systematic studies of the signal of guided molecules also for very small reservoir pressures. Thereby it is possible to confirm that the signal of the QMS is indeed proportional to the density of the guided molecules in the ionization volume and not to its flux. In particular, saturation of the ionization yields does not occur for the molecular speeds occurring in these measurements.

\section{Velocity distributions of guided molecules}
\label{sec:VelDist}

\begin{figure}
\centering
\includegraphics[width=.6\textwidth]{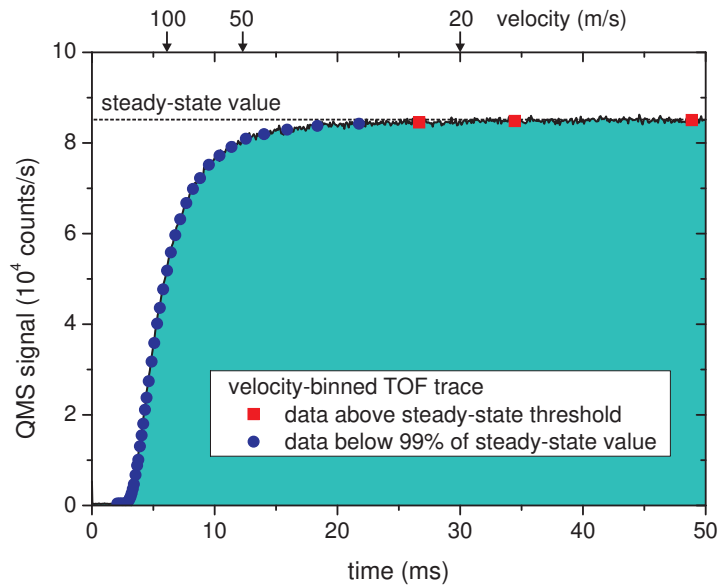}
\caption{\label{fig:TOFtrace}Time of flight trace of guided molecules. After switching on the high-voltage on the guide electrodes (at $t$=0), the signal rises and finally reaches a steady-state value. To derive a velocity distribution, velocity-dependent count rates are calculated from the arrival times and binned with a resolution of 5\,m/s (blue points). To avoid systematic effects, data above a given threshold fraction of the steady-state value is dropped (red squares).}
\end{figure}

To obtain additional information on the guided beam of polar molecules, a velocity distribution can be constructed from the time-resolved QMS signal. For that purpose, the arrival-time distribution of the molecules is analyzed. As shown in \fref{fig:TOFtrace}, after switching on the high-voltage the QMS signal rises to a steady-state value which was used for the measurements discussed so far. Using this time-of-flight (TOF) signal, the velocity of the molecules is calculated from the arrival times. For a given time $t$ after switching on the guiding fields, molecules with velocities $v\geq d/t$ contribute to the signal, where $d$ is the total length of the electric guide. Then, a binning of these velocities with a bin width of typically 5\,m/s is applied. This value is chosen as a compromise between resolution and sufficient signal-to-noise ratio. In a next step, a post selection  is applied to our data. As can be seen from the TOF trace in \fref{fig:TOFtrace}, the signal has already closely approached the steady-state value for late arrival times, corresponding to low velocities. Therefore, these data might be more affected by systematic errors, since a small effect on the count rate could already lead to a big difference in the velocity distribution. To avoid systematic effects, all the data above a given threshold, typically 99\,\%, of the steady-state value, is not considered in this evaluation. This can lead to a variation of the lowest velocity data point. To finally derive a velocity distribution, differences in count rates between these bins are taken.

\begin{figure}
\centering
\subfloat[$\pm$3\,kV electrode voltage]{
\includegraphics[width=0.48\textwidth]{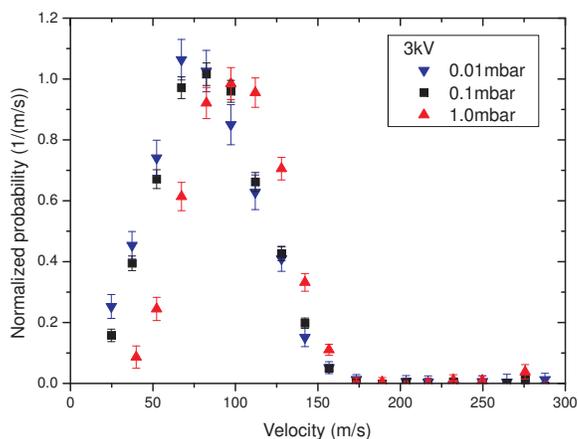}
}
\hfill
\subfloat[$\pm$5\,kV electrode voltage]{
\includegraphics[width=0.48\textwidth]{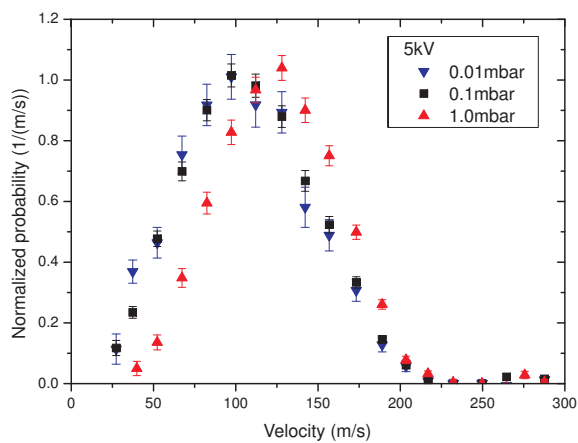}
}
\caption{\label{fig:VelDist}Normalized velocity distributions for different reservoir pressures. The effect of collisions for increasing reservoir pressure is visible as a shift of the velocity distribution towards higher velocities, while the cut-off velocity stays the same.}
\end{figure}

The velocity distributions give us another, even more direct, method to look at the boosting feature. Therefore, to see effects of collisions, we compare velocity distributions determined for fixed electrode voltage but for different reservoir pressures. The effects of boosting on the guided beam are visible in their clearest form by the dependence of the velocity distributions on the reservoir pressure. \Fref{fig:VelDist} shows normalized velocity distributions obtained from TOF traces for different reservoir pressures. The velocity distribution gets shifted to higher velocities for increasing reservoir pressures. As expected, the cut-off velocity stays the same. The cut-off velocity on the high velocity side depends only on the dominant Stark shift of the guided molecules and the guide geometry. Molecules with a velocity exceeding this cut-off velocity cannot be guided. This also explains the shape of the pressure dependence curve discussed in \Sref{sec:PScan}: For increasing reservoir pressure, larger and larger parts of the velocity distribution of molecules introduced into the guide is shifted beyond the cut-off velocity. Thereby, the fraction of molecules with velocities accepted by the guide decreases.

By comparing the velocity distributions for $\pm$3\,kV and $\pm$5\,kV electrode voltage in \fref{fig:VelDist}~(a) and \ref{fig:VelDist}~(b), one observes a shift in the cut-off velocity as well as in the position of the maximum of the distribution. The reason for this is the increase of the trapping field with electrode voltage. At larger trapping fields a larger velocity range of molecules is accepted by the guide as already discussed in \Sref{sec:VelFilter}.

If one is interested only in the slowest molecules from the velocity distribution, increasing the pressure for an increase in guided signal is counterproductive. As discussed in \Sref{sec:PScan}, the pressure for the optimum signal of guided molecules shifts to smaller values with reduced electrode voltage. This can also be observed from the velocity distribution. For 1.0\,mbar reservoir pressure, for instance, the slowest molecules observed in the experiment have velocities of around 35--40\,m/s, while for 0.1\,mbar reservoir pressure molecules down to 25\,m/s are observed. Especially in the $\pm$3\,kV measurement it can be seen that by decreasing the pressure from 0.1\,mbar further to 0.01\,mbar, the rising slope of the velocity distribution shifts to lower velocities. This is in agreement with the voltage dependencies shown in \Sref{sec:VDep}, which also show an increase in guided signal for low voltages, corresponding to slow molecules, when the pressure is reduced that way. Even for this very small reservoir pressure, extrapolation of the rising slope of the velocity distribution does not cut the velocity axis at zero velocity. Also the electrode voltage dependence measurements for low (0.01\,mbar) reservoir pressure show that for small electrode voltages less molecules are detected than expected from the $U^{3/2}$ power law valid for a purely effusive source. This indicates  that even for such reduced pressures collisions still play a role for the slowest molecules coming out of the nozzle. In the $\pm$5\,kV measurement this effect is less visible. This can be attributed to the very limited sensitivity of the TOF method on the low velocity side. For the lowest velocity in the distribution, the TOF signal has already reached nearly 99\% of the steady state value. Therefore, the rising slope of the velocity distribution might be more susceptible to systematic errors even after rejecting the long-time data.

For many applications, however, the boosting should not be a problem. When the reservoir pressure is increased, the maximum of the velocity distribution does not shift too much: For an increase of the reservoir pressure from 0.1\,mbar to 1.0\,mbar the shift is only $\approx$25\,m/s at $\approx$100\,m/s peak velocity at $\pm$5\,kV. Nonetheless, this raise of the reservoir pressure increases the flux of guided molecules and thereby the detector signal by a factor five. This makes the source very attractive for, e.g., spectroscopic studies where a high density of cold molecules within a certain velocity interval is desirable.

\section{Conclusion and Outlook}
\label{sec:Outlook}

To summarize, we have studied the performance of the electric guide over a wide parameter range. This allows to verify a model for the beam formation taking into account collisions in and near the nozzle. In comparison with an ideal effusive source, these collisions reduce the fraction of slow molecules and thereby the extraction efficiency. By operating in a dedicated pressure range, however, an optimized flux of molecules up to a given maximum velocity can be achieved. The effects of collisions in the formation of the beam are directly observed in the velocity distributions as well as in the characteristic shape of the electrode voltage dependence of the signal of guided molecules. The results are in agreement with our experiments employing a cryogenic helium buffer-gas cell for cooling of the molecules, where the source performance is limited only by the boosting for increasing helium densities \cite{vanBuuren2008,Sommer2008}. Already now, our beams are well suitable for collision experiments \cite{Willitsch2008} or spectroscopic applications, since they combine a high continuous flux with a relatively high purity \cite{Rieger2006,Motsch2007,Motsch2008a}. The detailed understanding of the electric guide as a source for cold polar molecules enables more insight for the design of future experiments, e.g., for an electric trap for molecules. There, further cooling of the molecules, e.g.\ by collision with ultracold atoms, should be possible when sufficiently long storage times and large enough densities can be reached. Also the development of a new source dedicated to experiments with only the slowest molecules benefits from these results.  The electric guiding technique offers many prospects for future experiments by providing a starting point for new cooling schemes.

\ack
Support by the Deutsche Forschungsgemeinschaft through the excellence cluster "Munich Centre for Advanced Photonics" and EuroQUAM (Cavity-Mediated Molecular Cooling) is acknowledged.

\end{document}